\documentclass[11pt,letterpaper]{article}
\usepackage[english]{babel}
\usepackage{fullpage}
\usepackage[margin=1in]{geometry}
\usepackage{graphicx}
\usepackage[utf8]{inputenc}
\usepackage{johd}
\usepackage{setspace}
\usepackage{url}
\title{Predicting Digital Asset Prices \\
        using Natural Language Processing: a survey}

\author{Trang Tran \\
        \small Cornell University, USA \\
        \small \tt{tmt59@cornell.edu}
}

\date{} 

\begin{document}

\maketitle

\begin{abstract} 
\noindent The introduction of blockchain technology has changed the way people think about how they used to store and trade their assets, as it introduced us to a whole new way to transact: using digital currencies. One of the major innovations of blockchain technology is decentralization, meaning that traditional financial intermediaries, such as asset-backed security issuers and banks, are eliminated in the process. Even though the blockchain technology has been utilized in a wide range of industries, its most prominent application is still cryptocurrencies, with Bitcoin being the first one proposed. At its peak in 2021, the market cap for Bitcoin once surpassed 1 trillion US dollars. The open nature of the crypto market poses various challenges and concerns for both potential retail investors and institutional investors, as the price of the investment is highly volatile and its fluctuations are unpredictable. The rise of Machine Learning, and Natural Language Processing, in particular, has shed some light on monitoring and predicting the price behaviors of cryptocurrencies. This paper aims to review and analyze the recent efforts in applying Machine Learning and Natural Language Processing methods to predict the prices and analyze the behaviors of digital assets such as Bitcoin and Ethereum.
\end{abstract}

\noindent

\section{Introduction} \label{introduction}
In recent years, blockchain technology and its desirable characteristics, such as decentralization and anonymity, have motivated the industry and academia. Even though the concept of using cryptographically secured chain of blocks was first described by \citet{haber1990time} in their paper "How to Time-Stamp a Digital Document", the term blockchain started to attract significant attention after the group of developers who go by the pseudonym Satoshi Nakamoto described the blockchain model their white paper \citep{nakamoto2008bitcoin} and implemented the peer-to-peer electronic cash system in 2009. The open and anonymous nature of the market has facilitated new opportunities while introducing new risks never seen before with established financial institutions, such as trust, unpredicted fluctuations, and fraudulent crowdfunding schemes.

The digital age also sees the rise of social media platforms such as Twitter and Reddit, where users can easily and conveniently express their opinions at any time and from anywhere, in the form of a simple post or a "tweet." Investment advice and opinions are among the most popular topics discussed online because of its real-time nature. While Twitter has implemented some forms of verifications to encourage people to use their real identities, such as asking for phone numbers or emails, Reddit has done little to ensure the users' identities are traceable. As a result, the world has witnessed the infamous case of Reddit's subforum "WallstreeBets" in January 2021, where the explosive power of Internet users and their speculative investment research can manipulate the investment prices, leading to the short squeeze of Gamestop's stock (GME) \citep{boylston2021wallstreetbets}. In an open space with less regulation and more freedom like cryptocurrency, we see a clear opportunity to utilize textual data from social media posts helping investors make informed investment decisions.

The remainder of this paper is organized as follows. In Section \ref{blockchain-background}, we briefly covered the background and terminologies behind blockchain. In Section \ref{nlp-background}, we similarly presented a brief literature review of the recent developments in the Natural Language Processing space. We dived deep into the recent works that use Natural Language Processing to predict the price and behaviours of cryptocurrencies in Section \ref{applications}, and concluded the paper in Section \ref{conclusion}.

\section{Background on Blockchain technology} \label{blockchain-background}
At a high level, a blockchain is a distributed ledger where all transactions are recorded in a chain of blocks. As new transactions are committed, new blocks will be appended to the chain \citep{zheng2018blockchain}. A distributed ledger is a public database where all participants can contribute and maintains a synchronized copy of the data.

\subsection{Key characteristics}
\begin{itemize}
    \item \emph{Decentralization.} A key innovation of blockchain technology is the elimination of traditional intermediaries (central banks). Cost and flow time can be massively reduced since transactions can happen peer-to-peer (P2P). There are no single points of failure, as every network participant has the same copy of the data.
    \item \emph{Persistency.} blockchain is built to be an immutable ledger, meaning it is impossible to alter or delete the data already recorded in the blocks. 
    \item \emph{Anonymity.} In the blockchain network, users are not identified by real names but by a randomly generated address, which will be liked to each user's transaction. Even though the transactions and the user's address (or wallet address) are public, the keys to prove ownership is private to each user.
    \item \emph{Auditability.} Since all participants have a copy of the data; it is possible to verify and trace the transactions based on the public wallet address and the digital timestamp.
\end{itemize}

\subsection{Types of blockchain network}
There are three major types of blockchain networks: Public, Private and Hybrid.
\begin{itemize}
    \item \emph{Public (Permissionless) blockchain.} In this network, users can join and leave at any time without obtaining prior permission.
    \item \emph{Private (Permissioned) blockchain.} In this network, only a limited number of users with permissions can read and write.
    \item \emph{Hybrid blockchain.} This is a combination of both public and private blockchains. Transactions and private information are kept inside the network but can still be verified.
\end{itemize}

\subsection{Bitcoin (BTC) and Ether (ETC) and other crytocurrencies} 
Bitcoin, proposed by \citet{nakamoto2008bitcoin}, is a digital currency using blockchain technology to facilitate transactions. It is still the most popular general-purpose cryptocurrency and is considered the first cryptocurrency ever created. There are 21 million bitcoins in total, and each bitcoin consists of 100 million smaller units (satoshis). Bitcoin is an example of public blockchain.

Since Bitcoin launched, other alternative cryptocurrencies based on blockchain technology have been introduced, including Ethereum - developed by \citet{buterin2014}. Its focus is providing a protocol for building decentralized applications (dApps) deployed on Ethereum virtual machines (EVMs). Users can create smart contracts to perform various tasks using a built-in Turing-complete programming language to be executed on the blockchain. \emph{Turing complete} (or computationally universal) refers to the idea that a system needs to have the capacity to implement arbitrary computer algorithms. In other words, it needs to be able to stimulate a Turing machine. Ether (ETH) is the native cryptocurrency on the Ethereum network. Other popular cryptocurrencies by market capitalization include Tether (USDT), USD Coin (USDC), Binance Coin (BNB), etc. Both Tether, USD Coin, and Binance Coin were initially built on the Ethereum network.

\subsection{Predicting Prices and Trends of Cryptocurrency}
As cryptocurrencies are considered valuable digital assets, various studies have been done to analyze and protect cryptocurrencies, just like any other type of investment. Past works have used Machine Learning methods extensively, alongside classical time-series variables (i.e., closing prices), to predict prices of cryptocurrencies such as Bitcoin, Ethereum, etc. Popular choices for model architecture include Autoregressive Integrated Moving
Average (ARIMA) \citep{garg2018autoregressive, roy2018bitcoin}, Support Vector Machine (SVM) \citep{poongodi2020prediction}.

In recent years, with the advancement in Machine Learning and Natural Language Processing, we have seen broad adoption of language models to support past works in predicting the Prices and Trends of an investment. Prior research suggested that peers' influence and public moods in the form of unstructured texts have been found to play an important role in investment decisions and market predictions \citep{farrell2022democratization, farimani2021leveraging, vo2019sentiment}. As a result, it is common to see an architecture in recent literature comprising two parts: a sentiment analyzer (to process the textual data such as Tweets from Twitter) and a prediction model (to predict the prices).

\begin{figure}
    \centering
    \includegraphics[width=0.5\textwidth]{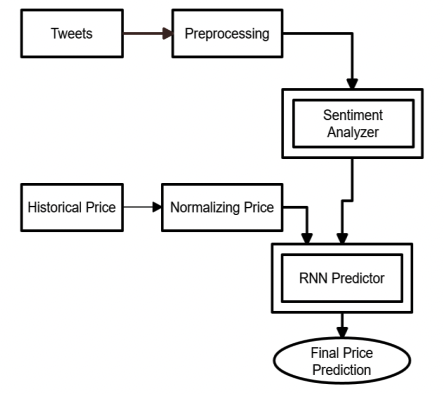}
    \caption{An example system to predict the price of bitcoin by \citet{pant2018recurrent}. The Recurrent Neural Networks (Section \ref{sec:rnn}) take into account two inputs: a sentiment score from a tweet, and the historical price of bitcoin to predict the next price}
    \label{fig:typical_architecture}
\end{figure}

\section{Background on Natural Language Processing} \label{nlp-background}
To tackle a large amount of textual data, earlier works in the area have focused on gauging the general sentiment or the polarity of the text and quantifying it using manual labels (i.e., "positive," "neutral," and "negative" that are associated with a numeric value). This is generally considered the lexicon-based approach, in which a sentence will be transformed into a vector via a simple method such as a token count matrix \citep{csacsmaz2021tweet}. After that, a simple classification model such as Logistic Regression, Support Vector Machine (SVM), or Naive Bayes classifier. Latest developments in the Natural Language Processing field have pushed the frontier in improving the performance of sentiment classification system by allowing the vector representation of a word to have a context. Various recent state-of-the-arts language models introduced recently include BERT \citep{devlin2018bert}, RoBERTa \citep{liu2019roberta}, XLNet \citep{yang2019xlnet}, and ELECTRA \citep{clark2020electra}.

\subsection{Artificial Neural Networks (ANNs)}
\begin{figure}
    \centering
    \includegraphics[width=\textwidth]{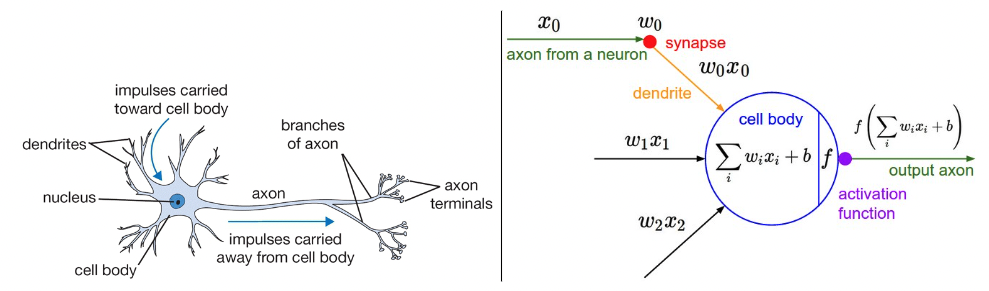}
    \caption[Caption]{A biological neuron (left) and its simplified mathematical model (right)\protect\footnotemark}
    \label{fig:ann}
\end{figure}

\begin{figure}
    \centering
    \includegraphics[width=0.5\textwidth]{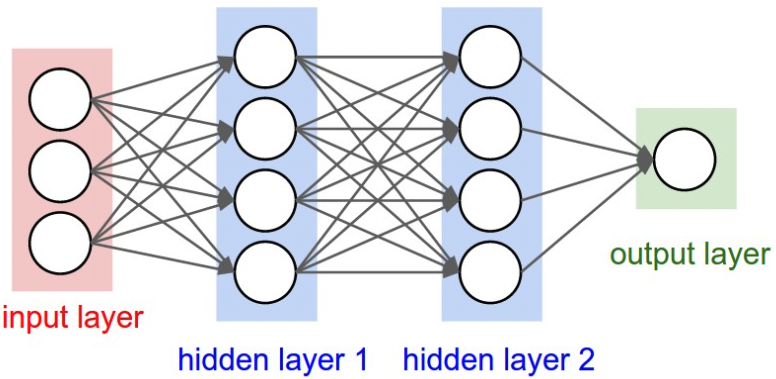}
    \caption[Caption]{Example architecture of a simple 3-layer neural network\protect\footnotemark}
    \label{fig:ann_architecture}
\end{figure}

An artificial neural network, or a neural network for short, is a mathematical model primarily inspired by the structure and functionalities of the biological neural system in the brain. Figure \ref{fig:ann} showed the connection between a biological neuron on the left and its mathematical model on the right. At a high level, they work by taking input signals from their dendrites and output signals along its axon. These axons eventually connect to the dendrites of other neurons, creating the entire \emph{network}. Another name for ANNs is Feed-Forward Neural network because the input information will only be processed in the forward direction, as seen in Figure \ref{fig:ann_architecture}.

\footnotetext{\url{https://cs231n.github.io/neural-networks-1/}}

\footnotetext{\url{https://cs231n.github.io/neural-networks-1/}}

\subsection{Recurrent Neural Networks (RNNs)} \label{sec:rnn}
Recurrent Neural Networks (RNNs) are models for data that involve times or sequences, such as sound, video, or language. They addressed a few limitations of Feed-Forward Neural networks, such as keeping track of or "memorizing" some of the inputs to make predictions in a situation where context and historical information plays an important role. 

\subsection{Long short-term memory Networks (LSTMs)}
Long Short-Term Memory networks or LSTMs are a type of RNN introduced by \cite{hochreiter1997long} to address the long-term dependency problem in the original RNN architecture. The original RNN cannot make an accurate prediction as the gap between the point containing the relevant information and the point where the relevant information increases. As a result, LSTMs have become the popular model architecture with significant use cases in Image captioning, Language Translation, Algorithmic trading, and Text Generation.

\section{Applications of NLPs in price prediction} \label{applications}
We can categorize the works in this space into two major groups: Corpus-based, and Advanced methods.  

\subsection{Corpus-based methods}
Corpus-based methods are based on the actual, real, and authentic words as they showed up in the document \citep{almutairi2016effectiveness}. By comparing and contrasting raw words from a document with a dictionary like the Harvard-IV dictionary (Dictionary with a list of positive and negative words), researchers can quickly sense the overarching sentiment of a post by aggregating the number of times that a positive or negative word shows up. \citet{sul2017trading} utilized this method to show that a trading strategy incorporating Twitter postings' sentiments could generate an annual return between 10 and 15\%.

\subsection{Advanced NLP methods}

Advanced NLP techniques attempt to go beyond relying purely on the count of the words as they occur in a sentence. These methods gears towards language understanding, as they allow the context to be incorporated into the vector representation of a word, as discussed in Section \ref{nlp-background}.

\begin{figure}
    \centering
    \includegraphics[width=0.66\textwidth]{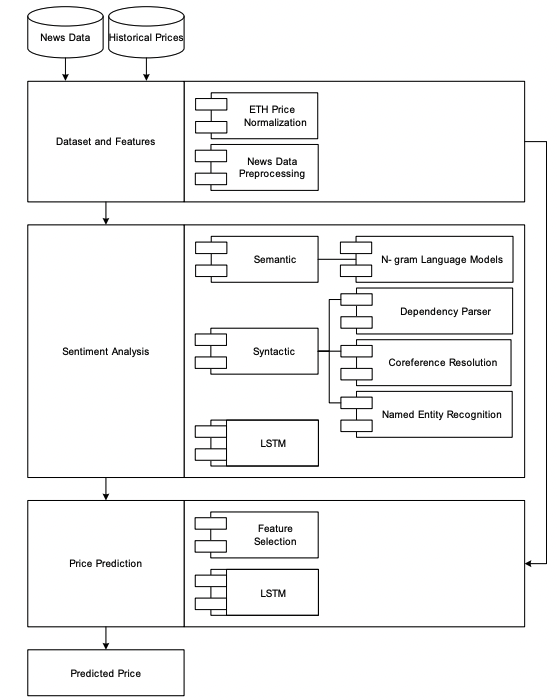}
    \caption{Proposed system to predict the price of Ethereum
    \citep{vo2019sentiment}}
    \label{fig:vo_overall_architecture}
\end{figure}

\begin{figure}
    \centering
    \includegraphics[width=0.68\textwidth]{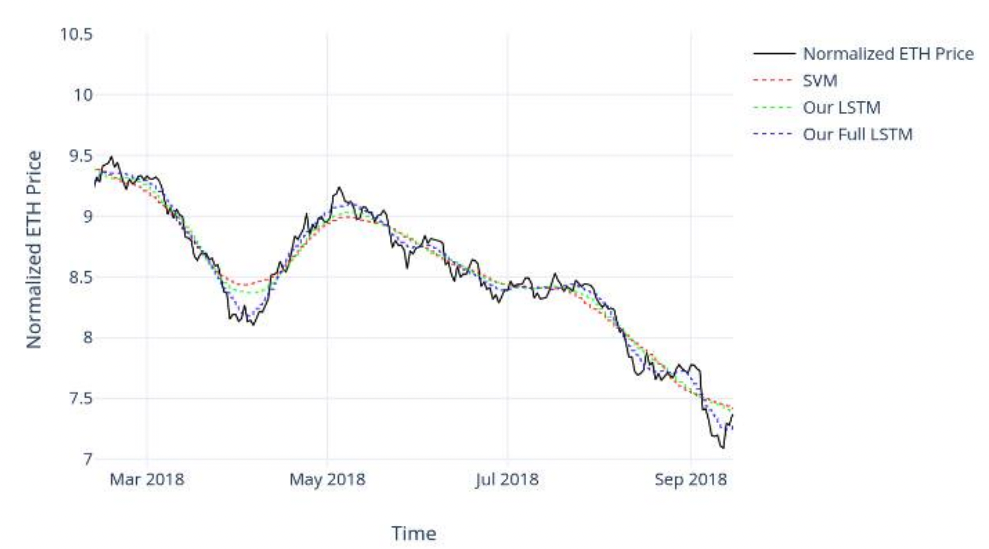}
    \caption{Result of predicting Ethereum (ETH) prices using different approaches \citep{vo2019sentiment}}
    \label{fig:vo_sentiment_price}
\end{figure}

\citet{vo2019sentiment} built a price prediction system using both news data and historical prices of Ethereum (ETH) to generate a prediction for future prices (Figure \ref{fig:vo_overall_architecture}). The authors created semantic vectors of the news document using a classic $n$-gram language model and syntactic vectors utilizing a set of NLP methods such as Dependency Parser, Coreference Resolution, and Named Entity Recognition. These two vectors are combined to predict the sentiment score. Eventually, the sentiment score and the historical cryptocurrency price are used to predict the future price. The text data used in this work is crypto-related, obtained from NewsNow\footnote{\url{https://newsnow.co.uk}}, and manually labeled. A label of 1 indicates a positive sentiment before a rise in the historical price and vice versa. Otherwise, the news will be labeled as 0 (neutral) as there is no change in the price. Figure \ref{fig:vo_sentiment_price} showed that the proposed architecture achieved relatively better performance than the SVM baseline when predicting ETH prices between July 2017 and October 2018.

To leverage the power of transfer learning, \citet{cheuque2021bitcoin} used the pretrained ULMFiT and BERT model in predicting the stance of a tweet from crypto-influencers and used these insights in building price prediction models using XGBoost and LSTM. The author showed that while there might be a weak relationship between the asset price and crypto-influencers’ opinion, this effect was much stronger and more causal in the short term.

\begin{figure}
    \centering
    \includegraphics[width=\textwidth]{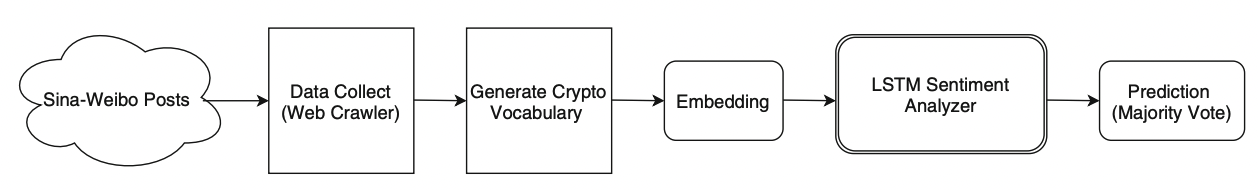}
    \caption{Cryptocurrency sentiment analysis and price movement prediction \citep{huang2021lstm}}
    \label{fig:huang_lstm}
\end{figure}

\citet{huang2021lstm} proposed using news articles and social media platforms such as Sina-Weibo, WeChat, and QQ groups to build a crypto-specific sentiment analysis model based on the Long short-term memory (LSTM) Recurrent Network Structure (Figure \ref{fig:huang_lstm}). The data used in the training process was manually labeled as positive (1), neutral (0), and negative (-1). The social media posts were tokenized using the manually-created crypto word vocabulary specific for Chinese words and used as the input to the LSTM Sentiment Analyzer. 

\citet{ortu2022cryptocurrency} showed via various empirical analyses that the price of Bitcoin is highly correlated with the sentiment from the discussions on social media forums such as Reddit. Using dynamic topic modeling and Hawkes models, they were able to explain how public opinions on social media chatter can influence each other and cause price fluctuations.

\citet{ortu2022technical} compared the performance of four different deep learning models (Multi-Layer Perceptron, Multivariate Attention Long Short Term Memory Fully Convolutional Network, Convolutional Neural Network, and Long Short Term Memory in predicting the price movements of Ethereum and Bitcoin between 2017 and 2020, utilizing BERT to extract the emotions from social media posts. The competitive results showed a lot of promise in detecting the price movements of Bitcoin and Ethereum.

\section{Conclusion} \label{conclusion}

The effort in using advanced machine methods to better understand cryptocurrencies has proved that investing in digital currencies is considered an alternative investment. As innovations are introduced in both the crypto world and the natural language processing field every day, opportunities and challenges exist in this niche research area. With cryptocurrency being a new research topic, more work needed to be done in this area to create an even more robust and accurate monitoring system to better utilize the massive amount of data available, especially the textual data surrounding the topic.

\bibliographystyle{johd}
\bibliography{bib}

\end{document}